\documentclass[aps,prl, superscriptaddress,twocolumn,groupedaddress,showpacs]{revtex4}
\usepackage{graphicx}
\usepackage{amsmath}
\begin{document}

\title{Size-dependent decoherence of excitonic states
in semiconductor microcrystallites}

\author{Yu-xi Liu}
 \affiliation{The Graduate University for Advanced Studies
 (SOKENDAI), Shonan Village, Hayama, Kanagawa 240-0193, Japan}
\author{Adam Miranowicz}
 \affiliation{The Graduate University for Advanced Studies
 (SOKENDAI), Shonan Village, Hayama, Kanagawa 240-0193, Japan}
 \affiliation{CREST Research Team for Interacting Carrier
 Electronics, Hayama, Kanagawa 240-0193, Japan}
 \affiliation{Nonlinear Optics Division, Physics Institute, Adam
 Mickiewicz University, 61-614 Pozna\'n, Poland}
\author{\c{S}ahin K. \"Ozdemir}
 \affiliation{The Graduate University for Advanced Studies
 (SOKENDAI), Shonan Village, Hayama, Kanagawa 240-0193, Japan}
 \affiliation{CREST Research Team for Interacting Carrier
 Electronics, Hayama, Kanagawa 240-0193, Japan}
\author{Masato Koashi}
 \affiliation{The Graduate University for Advanced Studies
 (SOKENDAI), Shonan Village, Hayama, Kanagawa 240-0193, Japan}
 \affiliation{CREST Research Team for Interacting Carrier
 Electronics, Hayama, Kanagawa 240-0193, Japan}
\author{Nobuyuki Imoto}
 \affiliation{The Graduate University for Advanced Studies
 (SOKENDAI), Shonan Village, Hayama, Kanagawa 240-0193, Japan}
 \affiliation{CREST Research Team for Interacting Carrier
 Electronics, Hayama, Kanagawa 240-0193, Japan}
 \affiliation{NTT Basic Research Laboratories, 3-1
 Morinosato-Wakamiya, Atsugi, Kanagawa 243-0198, Japan}

\date{\today}

\begin{abstract}
The size-dependent decoherence of the exciton states resulting
from the spontaneous emission is investigated in a semiconductor
spherical microcrystallite under condition $a_{B}\ll
R_{0}\leq\lambda$. In general, the larger size of the
microcrystallite corresponds to the shorter coherence time. If the
initial state is a superposition of two different excitonic
coherent states, the coherence time depends on both the overlap of
two excitonic coherent states and the size of the
microcrystallite. When the system with fixed size is initially in
the even or odd coherent states,  the larger average number of the
excitons corresponds to the faster decoherence. When  the average
number of the excitons is given, the bigger size of the
microcrystallite  corresponds to the faster decoherence. The
decoherence of the exciton states for the materials GaAs and CdS
is numerically studied by our theoretical analysis.


\end{abstract}

\maketitle \pagenumbering{arabic}

The field of quantum computation and information processing is
being extensively investigated due to possibility of finding
solutions to some intractable problems on classical computers
using the paradigm of quantum mechanics. Superposition principle
of quantum mechanics forms the strong basis for the realization of
the quantum computation and information processing schemes. The
carrier of the quantum information is called as qubit which can
take not only logical zero or one values as in its classical
counterpart, but also their superposition.  Any two-level system
is suitable for the realization of qubits.

As low-dimensional semiconductor structures, quantum dots are very
promising candidates for carrying quantum information because of
their atom-like properties.  The experimentalists have made great
progress in the coherent observation and manipulation of states in
 systems of quantum dots~\cite{st,jp,hirayama}, including demonstration
of the quantum entanglement of excitons in a single dot~\cite{5}
and a quantum dot molecule~\cite{6}, or Rabi oscillations of
excitons in single quantum dot~\cite{kam}. The quantum gate
realization based on the localized electron spins of quantum dots
as qubits has been proposed ~\cite{dot1}. A scheme of the
controllable interactions between two distant quantum dot spins by
combining the cavity quantum electrodynamics (QED) with electronic
spin degrees of freedom in quantum dots was
proposed~\cite{imamoglu}, and the degree of quantum entanglement
was further analyzed~\cite{Adam}. The reference~\cite{7}
theoretically investigates the entanglement of excitonic states in
the system of the optically driven coupled quantum dots
 and proposes a method to prepare maximally
entangled Bell and Greenberger-Horne-Zeilinger states.
 An all optical implementation of quantum
information processing with semiconductor macro atoms based on
excitons was also proposed~\cite{bb}.

It is well known that  pure quantum superposition states can
hardly survive for a long time. Because the interaction of any
system with the surrounding environment is unavoidable in a
physical realization, the energy of the superposed state of the
system is dissipated into environment or the relative phase of the
superposed components is disturbed by the environment. The
environment effect on the qubit defined by the exciton state has
been investigated~\cite{liu,fj,zanardi} in quantum dot with fixed
size. However, in general, one can expect that increasing the size
of a quantum dot results in stronger decoherence rates due to the
increased overlap between the system and the environment.

In this report, the size-dependent decoherence is discussed in
detail for a system of the semiconductor spherical
microcrystallite (SSM) whose radius $R_{0}$ is smaller than the
wavelength $\lambda$ of the relevant radiation field, but much
larger than the Bohr radius $a_{B}$ of exciton in bulk
semiconductor, i.e. $a_{B}\ll R_{0}\leq\lambda$. We assume that
the excitation density is so low that the average number of
excitons in the Bohr radius volume is no more than one. Thus the
interactions between the excitons can be neglected. For the
optically allowed lowest-energy excitons, the size-dependent
Hamiltonian can be given under the rotating wave approximation
as~\cite{han}.
\begin{eqnarray}
H&=&\hbar\Omega
b^{\dagger}b+\hbar\sum_{k}\omega_{k}a^{\dagger}_{k}a_{k}\nonumber\\
&+&\hbar\chi\sum_{k}\frac{\hat{e}_{k}\cdot\overrightarrow{\mu}_{cv}}{\sqrt{\hbar
c k}}(b^{\dagger}a_{k}+ba^{\dagger}_{k})\label{1}
\end{eqnarray}
with $\chi=\frac{4\Omega R^{3/2}_{0}}{\sqrt{V\pi a_{B}^{3}}}$.
Here $\hat{e}_{k}$ is  a unit polarization vector of radiation
field, $\overrightarrow{\mu}_{cv}$ is interband transition dipole
moment, $V$ is volume of the microcrystallite, $k$ is wave vector
of the radiation field, $a_{k}(a_{k}^{\dagger})$ are the
annihilation (creation) operators of radiation field with
frequency $\omega_{k}$, and $b(b^{\dagger})$ is the annihilation
(creation) operator of the exciton  with the transition energy
$\hbar\Omega=E_{g}-E^{b}_{exc}+\frac{\hbar^{2}\pi^2}{2MR^2_{0}}$
where $E_{g}$ is the energy gap between the conduction and valence
bands, $E^{b}_{exc}$ is binding energy of the excitons and  $M$ is
the mass for the center-of-mass motion which is a sum of the
effective masses $m_{e}$  and  $m_{h}$ for the electron and hole,
respectively,  in the conduction and valence bands, i.e.
$M=m_{e}+m_{h}$.  The operators for both radiation field and
excitons are bosons. The Hamiltonian (\ref{1}) shows that both the
exciton energy $\Omega$ and the coupling constants $
(\chi\hat{e}_{k}\cdot\overrightarrow{\mu}_{cv})/\sqrt{\hbar c k} $
between the exciton and radiation fields depend on the size
$R_{0}$ of the microcrystallite, which means that the different
sizes of the microcrystallites result in the different decay rates
of the excitonic states.

In order to show how the different sizes of microcrystallites
affect the coherence of the exciton states, we  give the
Heisenberg equations of motion for operators of the radiation
field and excitons corresponding to Hamiltonian (\ref{1})
as~\cite{scully,Yu-xi}
\begin{subequations}
\begin{eqnarray}
\frac{\partial B(t)}{\partial
t}=&-i\chi\sum_{k}\frac{\hat{e}_{k}\cdot\overrightarrow{\mu}_{cv}}{\sqrt{\hbar
c k}}A_{k}(t)e^{-i\omega_{k}t+i\Omega t}, &\label{2a}\\
\frac{\partial A_{k}(t)}{\partial
t}=&-i\chi\frac{\hat{e}_{k}\cdot\overrightarrow{\mu}_{cv}}{\sqrt{\hbar
c k}}B(t)e^{-i\Omega t+i\omega_{k}t},&
\end{eqnarray}
\end{subequations}
where we have applied rotating frame transformation to the exciton
and radiation field variables with $b(t)=B(t)e^{-i\Omega t}$ and
$a_{k}(t)=A_{k}(t)e^{-i\omega_{k} t}$. Then we can obtain the
formal solution of $A_{k}(t)$ as
\begin{equation}
A_{k}(t)=A_{k}(0)-i\chi\frac{\hat{e}_{k}\cdot\overrightarrow{\mu}_{cv}}{\sqrt{\hbar
c k}}\int_{0}^{t}B(t^{\prime})e^{-i\Omega
t^{\prime}+i\omega_{k}t^{\prime}}{\rm d} t^{\prime}\label{31}.
\end{equation}
Substituting $A_{k}(t)$ in Eq.(\ref{2a}), we get
\begin{eqnarray}
\frac{\partial B(t)}{\partial t}&
=&-i\chi\sum_{k}\frac{\hat{e}_{k}\cdot\overrightarrow{\mu}_{cv}}{\sqrt{\hbar
c k}}A_{k}(0)e^{-i\omega_{k}t+i\Omega
t}\nonumber\\
&-&\int_{0}^{t}B(t^{\prime}) K(t-t^\prime) d t^{\prime}\label{3}
\end{eqnarray}
with the time-dependent kernel function
\begin{equation}
K(t-t^{\prime})=\chi^{2}\sum_{k}\frac{|{\mu}_{cv}|^2\cos^2\theta}{\hbar
ck}e^{-i(\Omega-\omega_{k})(t^{\prime}-t)},
\end{equation}
where $\theta$ is the angle between the interband transition
dipole moment $\overrightarrow{\mu}_{cv}$ and electric field
polarization $\hat{e}_{k}$. Without loss of generality, we assume
that the interband transition dipole moment
$\overrightarrow{\mu}_{cv}$ of the optically allowed exciton is
along the $z$ axis and radiation is isotropic. In order to give
the time-dependent kernel function $K(t-t^{\prime})$, we change
the summation over $k$ into integral~\cite{han}
\begin{equation}
\sum_{k}\rightarrow\frac{2V}{(2\pi)^3}\int_{0}^{\infty}\int_{0}^{2\pi}\int_{0}^{\pi}\frac{\omega^{2}}{c^3}
d\omega \sin\theta d\theta d\varphi,
\end{equation}
where the integration is taken only  over the volume $V$ of the
microcrystallite, because the interband transition dipole moment
$\overrightarrow{\mu}_{cv}$ vanishes  outside   the
microcrystallite.  The assumption of $R_{0}\leq\lambda$ makes the
polariton effect negligible implying the rapid radiation of
excitons due to the breakdown of the translation symmetry.
Consequently, we can consider that the radiation light intensity
of  microcrystallite is going to be centered at the transition
frequency $\Omega$, which is known as the Weisskopf-Wigner
approximation~\cite{scully}. Then the kernel function is found as
\begin{eqnarray}
K(t-t^{\prime}) &=&32\pi
\left(\frac{R_{0}}{a_{B}}\right)^{3}\gamma_{s}
\delta(t^{\prime}-t)\label{6}
\end{eqnarray}
 with
$\gamma_{s}=\frac{4|{\mu}_{cv}|^2\Omega^3}{3\hbar (2\pi c)^3}$
depending on the size of microcrystallite. Replacing the kernel
function of Eq.(\ref{6}) in Eq. (\ref{3}) and taking the Laplace
transform, the exciton operator $B(t)$ is found as
\begin{equation}
B(t)=B(0)e^{-32\pi
(\frac{R_{0}}{a_{B}})^3\gamma_{s}t}-i\sum_{k}u_{k}(t)A_{k}(0),\label{8}
\end{equation}
 $u_{k}(t)$  is given as
\begin{equation}
u_{k}(t)=\frac{\chi \hat{e}_{k}\cdot \overrightarrow{\mu
}_{cv}}{\sqrt{\hbar ck}}\frac{\exp (-64\pi
\frac{R_{0}^{3}}{a^{3}}\gamma _{s}t)-\exp [-i(\omega _{k}-\Omega
)t]}{-64\pi \frac{R_{0}^{3}}{a^{3}}\gamma _{s}+i(\omega
_{k}-\Omega )}.
\end{equation}
From Eq.(\ref{8}), it can be found that the spontaneous emission
rate of exciton is $64\pi(R_{0}/a_{B})^3$ times higher than that
of an atom which has the transition frequency $\Omega$ and the
transition dipole moment $\overrightarrow{\mu}_{cv}$~\cite{han}.
This increase in the spontaneous emission rate can be explained by
the fact that the exciton is coherently excited over the whole
quantum microcrystallite. Thus coherent excitation results in a
coherent transition dipole moment and a superradiant character. It
is obvious that  the information, which is carried by the state of
the exciton, is lost with the time evolution because of the energy
dissipation due to spontaneous emission of the exciton. The loss
rate of the information depends on the factor $32\pi
\left({R_{0}}/{a_{B}}\right)^{3}\gamma_{s}$.

Now lets analyze how the size-dependent decoherence affect a qubit
defined in the computational basis of the vacuum state $|0\rangle$
and single-exciton state $|1\rangle$. If the exciton system is
initially in a qubit state $\alpha|0\rangle+\beta|1\rangle$ with
$|\alpha|^2+|\beta|^2=1$, and the radiation field is initially in
the multimode vacuum state $\prod_{k}|0\rangle_{k}$, after an
evolution time of $t$, the state of the system will
become~\cite{liu},

\begin{eqnarray}
&&|\psi(t)\rangle=U(t)[\alpha|0\rangle+\beta|1\rangle]\otimes\prod_{k}|0\rangle_{k}\nonumber\\
&&=\alpha|0\rangle\otimes\prod_{k}|0\rangle_{k}+\beta U(t)b^{\dagger}|0\rangle\otimes\prod_{k}|0\rangle_{k}\nonumber\\
&&=[\alpha|0\rangle +\beta e^{\{-32\pi
(\frac{R_{0}}{a_{B}})^3\gamma_{s}t-i\Omega t\}}|1\rangle]\otimes\prod_{k}|0\rangle_{k}\nonumber\\
&&+
\beta|0\rangle\otimes\sum_{k^{\prime}}u_{k^{\prime}}(t)e^{-i(\Omega
+\omega_{k^{\prime}})t}|1\rangle_{k^{\prime}}\otimes\prod_{k\neq
k^{\prime}}|0\rangle_{k},
\end{eqnarray}
where the properties of the time evolution operator
$U^{\dagger}(t)bU(t)=b(t)$ and $U(t)|0\rangle=0$ with
$U(t)=e^{-iHt/\hbar}$ are applied. Because we are interested in
the exciton system, the multimode radiation field degrees of
freedom are traced out and the reduced density operator of the
exciton system is found as
\begin{eqnarray}
\rho(t)&=&(1-\eta(t)|\beta|^2)|0\rangle\langle
0|+\alpha\beta^{*}F(t)e^{i\Omega t}|0\rangle\langle
1|\nonumber\\
&+&\alpha^{*}\beta F(t)e^{-i\Omega t}|1\rangle\langle
0|+\eta(t)|\beta|^2|1\rangle\langle 1|\label{eq:11}
\end{eqnarray}
with $\eta(t)=\exp\left\{-64\pi
(\frac{R_{0}}{a_{B}})^3\gamma_{s}t\right\}$.  The slowly varying
time-dependent factor $F(t)$ of the off-diagonal elements in
Eq.(\ref{eq:11}), which can be written as
\begin{equation}\label{eq:12}
F(t)=\exp\left\{-32\pi
(\frac{R_{0}}{a_{B}})^3\gamma_{s}t\right\}=\exp{(-t/\tau)},
\end{equation}
is used to characterize the coherence of the superposed excitonic
state $\alpha|0\rangle+\beta|1\rangle$. The characteristic time
$\tau$ of the decoherence is given as,
\begin{equation}
\tau=\frac{3\hbar^4 \pi^2 c^3 a_{B}^3M^3R^3_{0}}{2|\mu_{cv}|^2
[2MR^2_{0}(E_{g}-E^{b}_{exc})+\hbar^2\pi^2]^3}\label{eq:13}.
\end{equation}

\begin{figure}
\includegraphics[width=6.5cm]{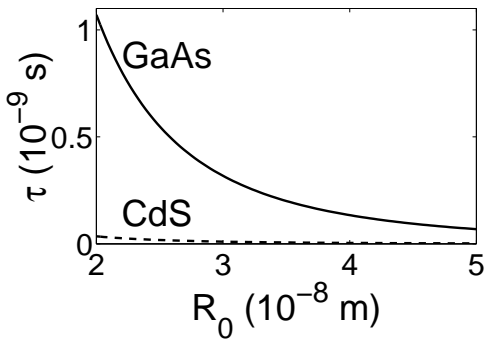}
\caption[]{Characteristic time $\tau$ as a function of size $200$
\AA~ $<R_{0}<500$ \AA~ for the CdS and GaAs microcrystallites when
the systems are initially in the superposition of the vacuum and
single-exciton states.}\label{fig1}
\end{figure}
To give  a better insight to the problem, we give some numerical
values for the chosen materials, CdS and GaAs. The following
characteristic parameters for those materials are taken from
references~\cite{han,tt}. For CdS, $E_{g}=2.583$ eV,
$E^{b}_{exc}=30 $ meV, $a_{B}=30$ \AA,
$\frac{|\mu_{cv}|^2}{(\epsilon_{0}a^3_{B})}=0.25$ meV with the
static dielectric constant of the bulk crystal $\epsilon_{0}=8$,
$m_{e}=0.25 m_{0}$ and $m_{h}=1.6 m_{0}$ where $m_{0}$ is the rest
mass of electron. For GaAs, $E_{g}=1.52$ meV, $E^{b}_{exc}=5$ meV,
$a_{B}=100$ \AA,
$\frac{|\mu_{cv}|^2}{(\epsilon_{0}a^3_{B})}=0.025$ meV with
$\epsilon_{0}=12.53$, $m_{e}=0.0665 m_{0}$ and $m_{h}=0.45 m_{0}$.
The area of the microcrystallite size is taken as $200$\AA $\leq
R_{0}\leq 500$\AA \ in   Fig. \ref{fig1}. In this range, the
wavelength is around $5000$ \AA \ for CdS, and $8000$ \AA \ for
GaAs, thus the condition  $a_{B}\ll R_{0}\leq\lambda$ is
satisfied.

In order to show how the characteristic time $\tau$ changes with
the size $R_{0}$, we give Fig. \ref{fig1}.  It is seen that a
larger size of the microcrystallite corresponds to a shorter
characteristic time $\tau$. So the larger is the size of
microcrystallite, the faster is the decoherence of the
superposition of two exciton states for a given material. But the
exciton state for different materials have the different
decoherence times even if the size of the microcrystallites are
the same. This is due to the different couplings of the systems
with the radiation field. Comparing the decoherence characteristic
time of CdS and GaAs in Fig. \ref{fig1}, we find that the
coherence time of the exciton state for  GaAs is longer than that
of CdS for a given size. This result shows the importance of
material choice in obtaining longer coherence time.

Here, we consider another kind of qubit state formed by the
superposition of coherent states, for example, we can define the
even coherent state ${\cal N}_{+}[|\alpha\rangle+|-\alpha\rangle]$
as the logic zero state $|0\rangle_{L}$ and the odd coherent state
${\cal N}_{-}[|\alpha\rangle-|-\alpha\rangle]$ as the logic one
state $|1\rangle_{L}$ where ${\cal N}_{\pm}=(2\pm
2e^{-2|\alpha|^2})^{-1/2}$ and $|\alpha\rangle$ is an eigenstate
of the bosonic operator. These states can be prepared by the
photon states in cavity quantum electrodynamics and by motional
states of trapped ions~\cite{haroche}.  Such logical qubit
encoding can be used as the correction of the spontaneous emission
errors~\cite{ptc}. We first assume that the exciton system is
initially in a superposition state of two coherent states
$|\alpha_{1}\rangle$ and $|\alpha_{2}\rangle$ of the exciton
annihilation operator
\begin{equation}
|\psi\rangle=C|\alpha_{1}\rangle+ D|\alpha_{2}\rangle,
\end{equation}
and the radiation field is in the multimode vacuum state
$\prod_{k}|0\rangle_{k}$. By using the same approach as
reference~\cite{Yu-xi} and Eq.(\ref{8}), we can find that the
time-dependent reduced density operator for the exciton system is
\begin{eqnarray}
\rho(t)&=&|C|^2|u(t)\alpha_{1}\rangle\langle u(t)\alpha_{1}|+
CD^{*}F(t)|u(t)\alpha_{1}\rangle\langle u(t)\alpha_{2}|\nonumber\\
&+& C^{*}D F^{*}(t)|u(t)\alpha_{2}\rangle\langle
u(t)\alpha_{1}|\nonumber\\
&+& |D|^2|u(t)\alpha_{2}\rangle\langle u(t)\alpha_{2}|
\end{eqnarray}
where $u(t)=\exp\{-32\pi
(\frac{R_{0}}{a_{B}})^3\gamma_{s}t-i\Omega t\}$ and the
time-dependent decoherence factor is
\begin{eqnarray}
F(t)&=&\exp\left\{\left[-\frac{|\alpha_{1}|^2}{2}-\frac{|\alpha_{2}|^2}{2}+
\alpha_{1}\alpha^{*}_{2}\right]\right.\nonumber\\
 &\times&\left.\left[1-\exp\left\{-32\pi
(\frac{R_{0}}{a_{B}})^3\gamma_{s}t\right\}\right] \right\}.
\end{eqnarray}
It is understood that the decoherence of the superposition state
of two coherent states is determined by both the properties of the
microcrystallite and the overlap between two coherent states. Due
to the existence of the vacuum state component in the excitonic
coherent state, it is also observed that the decoherence factor
$F(t)$ does not tend to zero in the long time limit $t \rightarrow
\infty$.

\begin{figure}
\includegraphics[width=40mm]{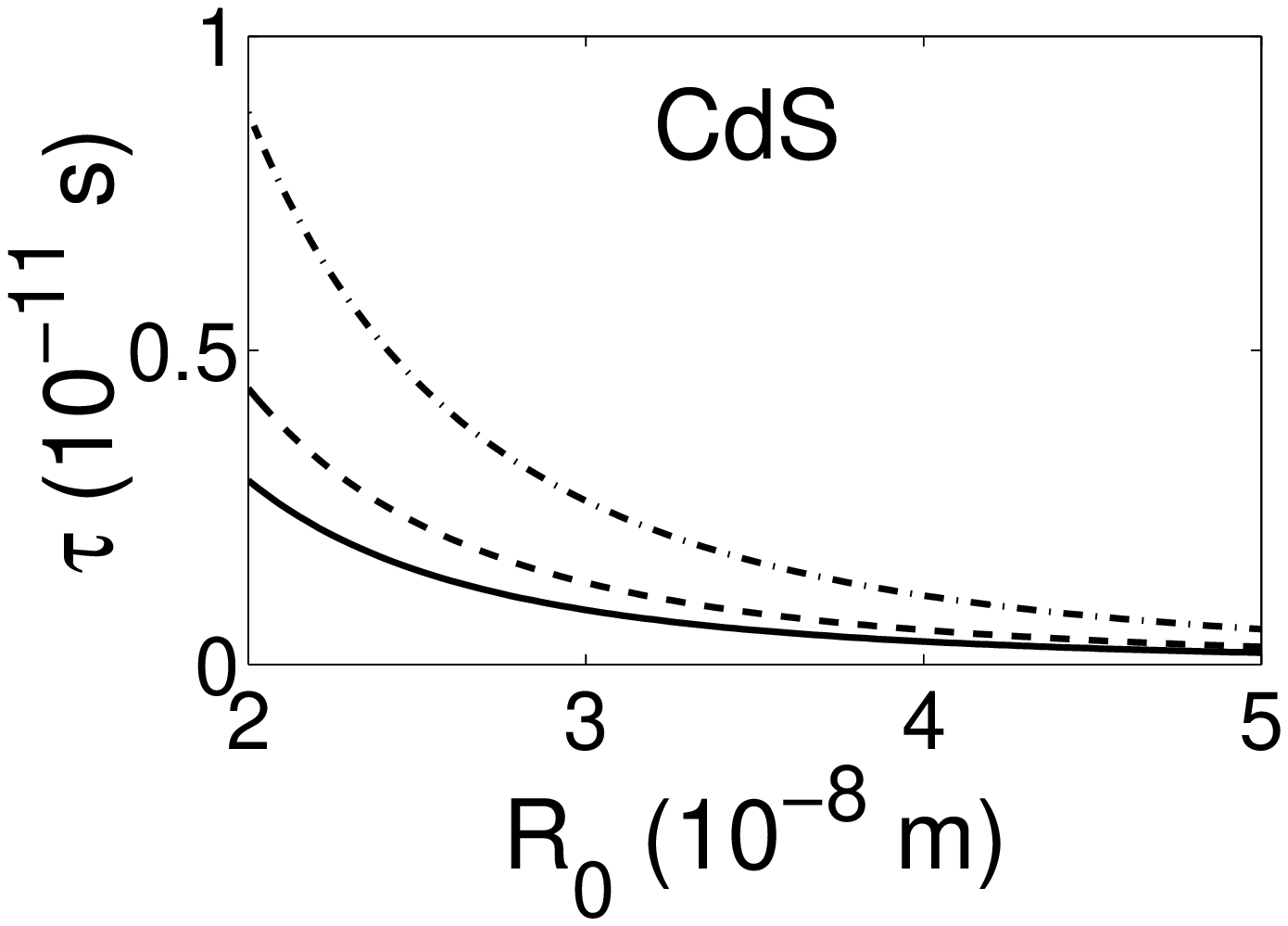}
\includegraphics[width=40mm]{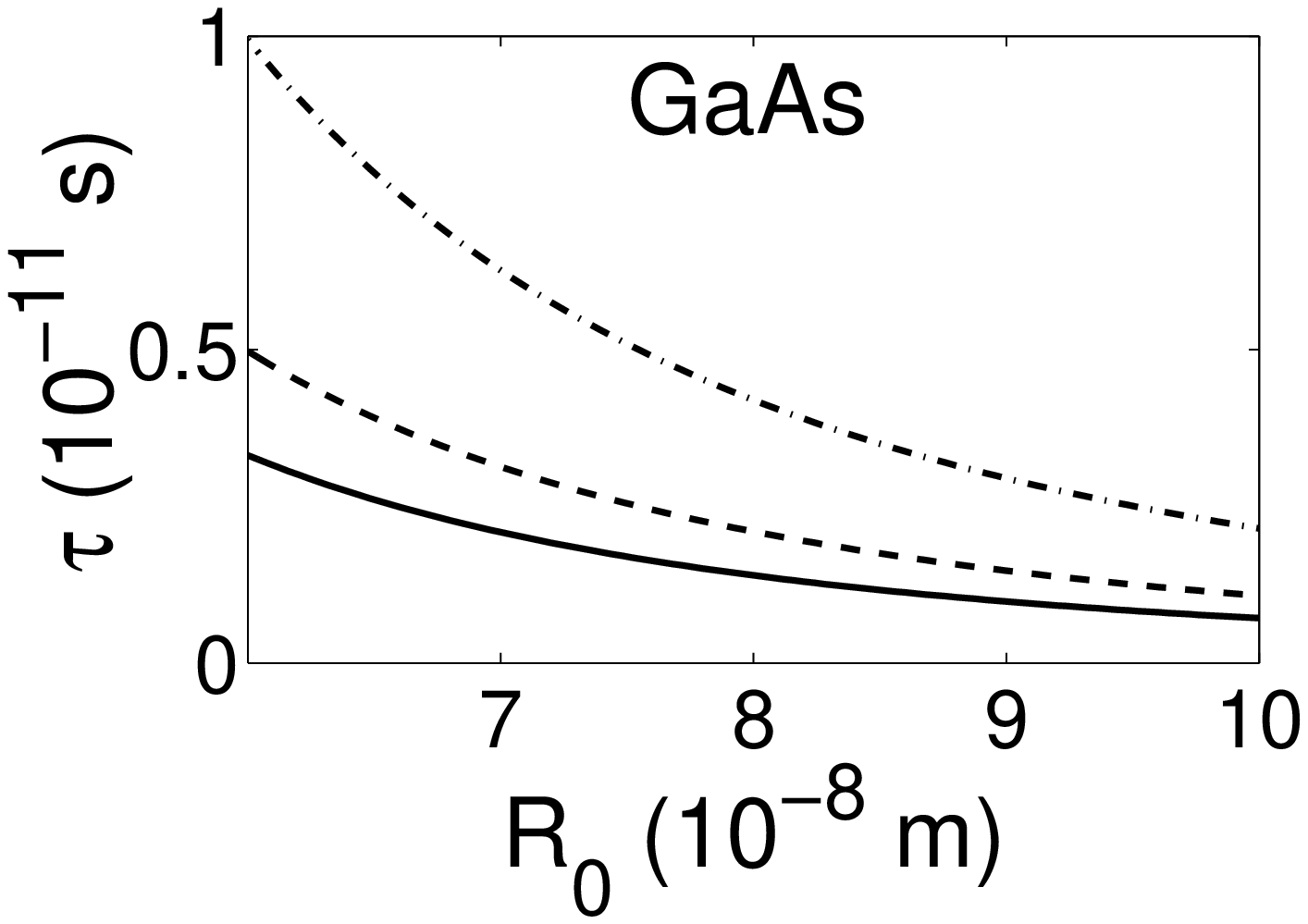}
\caption[]{Characteristic time $\tau$ as a function of size
$200$\AA~  $<R_{0}<500$ \AA~ for CdS and $600$\AA~  $<R_{0}<1000$
\AA~ for GaAs microcrystallites for different average numbers of
excitons: $|\alpha|^2=2$ (dot-dashed curves), $|\alpha|^2=4$
(dashed curves), $|\alpha|^2=6$ (solid curves)  when the systems
are initially in the even or odd coherent states}\label{fig2}
\end{figure}

For the qubit states which are defined by the even and odd
coherent states (also referred as the Schr\"{o}dinger cats), the
decoherence factor is
\begin{equation}
F(t)=\exp\left\{-2|\alpha|^2 \left[1-e^{\left\{-32\pi
(\frac{R_{0}}{a_{B}})^3\gamma_{s}t\right\}}\right]
\right\}\label{eq:17}.
\end{equation}
 For the
behavior of the short time $32\pi
(\frac{R_{0}}{a_{B}})^3\gamma_{s}t\ll 1$, the characteristic time
$\tau$ of the decoherence for the superposed excitonic coherent
states is approximately $1/(2|\alpha|^2)$ times of that in Eq.
(\ref{eq:13}). In order to discuss how the characteristic time
depends on the size and the
 average number of the excitons in detail. We plot Fig. \ref{fig2} with
different average numbers of the excitons in ranges $200$ \AA \
$\leq R_{0}\leq 500$ \AA \ for CdS and $600$ \AA \ $\leq R_{0}\leq
1000$ \AA \ for GaAs, in which the conditions of the bosonic
descriptions of excitons and $a_{B}\ll R_{0}\leq \lambda$ are
still valid. It is found that the larger average number of the
excitons corresponds to the shorter characteristic time when the
size of the microcrystallite is fixed, and the larger size of the
microcrystallite corresponds to the faster decoherence when the
average number of the excitons is fixed.

In summary, we have studied the size-dependent decoherence for the
semiconductor microcrystallites. It is found that
microcrystallites with larger sizes have shorter coherence time of
the exciton states. If  the system is initially in the
superposition of two different excitonic coherent states, the
coherence characteristic time depends on both the sizes of the
microcrystallites and the overlap between the two coherent states.
A numerical study of the decoherence of the exciton states for the
materials GaAs and CdS is carried out based on our theoretical
analysis. This numerical analysis clearly shows the importance of
material choice on the decoherence characteristics of
microcrystallites. It also should be pointed out that our study
cannot  simply  be generalized to the small quantum dot whose
effective radius $R_{0}$ is smaller than the Bohr radius $a_{B}$
in the bulk semiconductor when there exist many excitons in a
quantum dot. Because the Pauli principle prohibits the two or more
excitons from occupying the same energy state  in the small
quantum dot. The details need to be studied further.

Yu-xi Liu is supported by Japan Society for the Promotion of
Science (JSPS).

\end{document}